\documentclass[reprint,aps,prl,floatfix,amsmath,amssymb,longbibliography,notitlepage]{revtex4-2}
\usepackage{graphicx}
\usepackage{physics}
\usepackage{multirow}
\usepackage{tabularx,booktabs,ragged2e,array}     
\usepackage{booktabs}
\usepackage{ragged2e}   
\usepackage[table]{xcolor}

\usepackage[colorlinks, linkcolor= blue, citecolor = blue, urlcolor=blue]{hyperref}
\usepackage{pifont}
\usepackage{cancel}
\usepackage{comment}
\usepackage{mathrsfs}
\usepackage{bm}
\usepackage[normalem]{ulem}

\def\nn{\nonumber}
\def\bea{\begin{eqnarray}}
\def\eea{\end{eqnarray}}
\def\be{\begin{equation}}
\def\ee{\end{equation}}
\def\tc{\textcolor}
\def\e{\varepsilon}

\def\kb{{\bm k}}
\def\yes{\ding{51}}
\def\no{\ding{55}}
\definecolor{mygreen}{HTML}{E7FF86}

\begin{document}
\title{Longitudinal Nonreciprocal Charge Transport with Time Reversal Symmetry}

\author{Harsh Varshney}
\email{hvarshny@iitk.ac.in}
\affiliation{Department of Physics, Indian Institute of Technology, Kanpur-208016, India.}
\author{Amit Agarwal}
\email{amitag@iitk.ac.in}
\affiliation{Department of Physics, Indian Institute of Technology, Kanpur-208016, India.}

\begin{abstract} 

Longitudinal nonreciprocal charge transport is usually associated with broken time-reversal symmetry, either from magnetic order or an external magnetic field. Here, we show that it can also arise in nonmagnetic conductors preserving time-reversal symmetry through disorder-induced asymmetric scattering. Within a semiclassical Boltzmann theory, skew-scattering and side-jump processes generate a finite longitudinal current quadratic in the electric field. Our symmetry analysis identifies 42 point groups that allow this longitudinal nonreciprocal response. As a concrete example, gated Bernal-stacked bilayer graphene shows a gate-tunable nonreciprocal response with clear enhancement near its Lifshitz transition. These results identify disorder-driven asymmetric scattering as a route to bulk longitudinal nonreciprocal charge transport in crystalline conductors.
\end{abstract}
\maketitle

\tc{blue}{{\textit{Introduction:---}}} Nonreciprocal charge transport (NCT), the change in electrical response under current or field reversal, is a basic nonlinear transport effect in noncentrosymmetric systems~\cite{Rikken2001electrical, Pop2014electrical, Wakatsuki2017nonreciprocal, hoshino2018nonreciprocal, yuki2020nonreciprocal, Nagaosa2024nonreciprocal, Dash2025nonlinear, nagahama2025control, Soori2024decohorence}. In a two-terminal measurement, it appears directly as a longitudinal effect: opposite current directions give different resistances. Junctions and interfaces can show strong nonreciprocity, but a bulk longitudinal response in a crystal is much harder to realize because the second-order current must survive after the Brillouin-zone averaging and crystal-symmetry constraints while remaining measurable in a dissipative conductor~\cite{morimoto2018nonreciprocal, Tokura2018nonreciprocal, Nagaosa2024nonreciprocal}.

Most known routes to bulk longitudinal NCT rely on second-order transport in systems with broken inversion and broken time-reversal symmetry~\cite{Wakatsuki2017nonreciprocal, Ideue2017bulk, morimoto2018nonreciprocal, Tokura2018nonreciprocal, Yasuda2020large, Ando2020observation, Li2021nonreciprocal, Zhang2022controlled, Ye2022nonreciprocal, Wang2023quantum, Li2024observation, Nagaosa2024nonreciprocal, wang2024intrinsic, SurezRodrguez2025nonlinear, Nakamura2025nonreciprocal}. These include electrical magnetochiral anisotropy~\cite{Rikken2001electrical, Rikken2005magneto, Pop2014electrical, Meng2021interface, Jiang2024electrical}, nonlinear Drude terms~\cite{zhao2024general, Jiang2024electrical, chen2025magnetic}, intrinsic band-geometry effects~\cite{gao2014field, kamal2021intrinsic, Gao2023quantum, chen2025quantum, Jiang2024electrical}, and Lorentz-force-induced skew scattering~\cite{Xiao2026lorentz, lu2025lorentz}. In each case, the longitudinal response vanishes when time-reversal symmetry is preserved. Therefore, broken time-reversal symmetry through magnetic order or an external magnetic field, seemed to be essential for bulk longitudinal NCT. The basic open question is whether a nonmagnetic conductor can support a bulk longitudinal nonreciprocal response at all.

\begin{figure}[tbp]
    \centering
    \includegraphics[width = \linewidth]{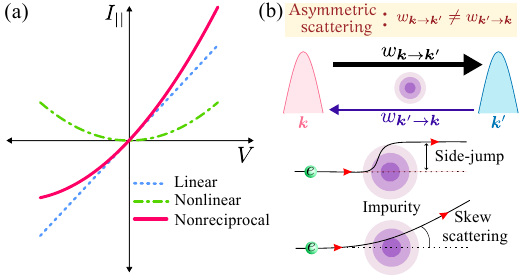}
    \medskip
    {\small
    \renewcommand{\arraystretch}{1.15}
    \setlength{\tabcolsep}{4pt}
    \begin{tabular*}{\linewidth}{@{\extracolsep{\fill}}lcc}
    \hline\hline 
    \textbf{Mechanisms} & $\cancel{\mathcal{T}}$ & $\mathcal{T}$ \\
    \hline\hline 
    Nonlinear Drude \cite{zhao2024general, wang2024intrinsic}  & \yes & \no \\ 
    Nonlinear Intrinsic~\cite{kamal2023intrinsic, chen2025magnetic} & \yes & \no \\ 
    Lorentz skew-scattering~\cite{Xiao2026lorentz, lu2025lorentz} & \yes & \no \\ 
    Nonlinear side-jump~\cite{du2019disorder, Datta2024nonlinear} & \yes & \cellcolor{mygreen} \yes \\ 
    Nonlinear skew-scattering~\cite{du2019disorder, Wang2025nonlinear}& \yes & \cellcolor{mygreen} \yes \\ 
    \hline
    \end{tabular*}
    }
    \caption{\textbf{Longitudinal nonreciprocal transport without magnetic fields.}
    (a) Schematic $I_{||}$–$V$ characteristics illustrating linear, nonlinear, and resulting nonreciprocal responses. A second-order contribution makes the forward and reverse currents unequal, producing a nonreciprocal current response.
(b) Microscopic mechanisms. In noncentrosymmetric systems,  asymmetric impurity scattering processes such as skew scattering and side jump generate a longitudinal current proportional to $E^2$, even when time-reversal symmetry ($\mathcal{T}$) is preserved. The table summarizes known mechanisms for longitudinal nonlinear transport and shows that extrinsic scattering processes enable nonreciprocity in nonmagnetic materials.
    \label{fig1}}
\end{figure}

In this Letter, we identify impurity-induced asymmetric scattering as a general mechanism for longitudinal NCT in noncentrosymmetric conductors preserving time-reversal symmetry (see Fig.~\ref{fig1}). The relevant extrinsic terms are nonlinear skew-scattering and side-jump contributions~\cite{Smit1955, Smit1958, Berger1970sidejump, du2019disorder, Ortix2021nonlinear, Wang2025nonlinear, SurezRodrguez2025nonlinear, ma2023anomalous}. They generate a longitudinal current proportional to $E^2$ even when time reversal is intact, while remaining consistent with microscopic reversibility, detailed balance, and Onsager constraints in a dissipative steady state (Appendix A1 and A2). A symmetry analysis across all 122 point groups identifies 42 crystal classes that permit this effect. As a concrete example, we demonstrate NCT in gated Bernal-stacked bilayer graphene, where a perpendicular electric field breaks inversion symmetry but preserves time reversal. We find a gate-tunable nonreciprocal response that is enhanced near the Lifshitz transition and is in qualitative agreement with recent experiments~\cite{Ahmed2025detecting, Huang2023intrisnic, panhe2022graphene}.

\tc{blue}{\textit{Origin of longitudinal nonlinear nonreciprocity:---}}
The charge current in a quasiparticle system is given by, 
$\bm j = -e \sum_{\bm k} \bm v_{\bm k} f_{\bm k}$, 
where $\bm v_{\bm k}$ is the band velocity and $f_{\bm k}$ is the distribution function. 
Under an applied electric field, both quantities acquire field-induced corrections. Expanding both quantities in powers of the electric field, we can express the second-order current as 
\begin{equation}
\bm j^{(2)} =
-e \sum_{\bm k}
\left[
\bm v^{(0)}_{\bm k} f^{(2)}_{\bm k}
+
\bm v^{(1)}_{\bm k} f^{(1)}_{\bm k}
+
\bm v^{(2)}_{\bm k} f^{(0)}_{\bm k}
\right]~.
\label{eq:j2_general}
\end{equation}
%
For a finite response proportional to $E^2$, the integrand must be even under momentum inversion $\bm k\!\rightarrow\!-\bm k$. In time-reversal-symmetric systems, the equilibrium distribution $f^{(0)}_{\bm k}$ is even in momentum while the band velocity $\bm v^{(0)}_{\bm k}$ is odd. Intrinsic electric-field corrections to the velocity, $\bm v^{(1)}_{\bm k}$ and $\bm v^{(2)}_{\bm k}$, are also odd in $\bm k$~\cite{sundram1999wavepacket,gao2014field}. Consequently the intrinsic contribution to $\bm v^{(2)}_{\bm k}f^{(0)}_{\bm k}$ remains odd and vanishes after Brillouin-zone integration.

The first-order nonequilibrium distribution~\cite{ashcroft1976ssp} satisfies $f^{(1)}_{\bm k}\!\propto\!\bm E\!\cdot\!\bm v^{(0)}_{\bm k}$ and is therefore odd in momentum. Although the product $\bm v^{(1)}_{\bm k}f^{(1)}_{\bm k}$ is even, the intrinsic velocity correction $\bm v^{(1)}_{\bm k}\!\propto\!\bm E\times\bm\Omega_{\bm k}$ is transverse to the applied field and produces nonlinear Hall responses rather than a longitudinal NCT. The remaining term $\bm v^{(0)}_{\bm k}f^{(2)}_{\bm k}$ can generate a longitudinal nonlinear current, but since $\bm v^{(0)}_{\bm k}$ is odd in momentum it requires a $\bm k$-odd component in $f^{(2)}_{\bm k}$. Within the conventional symmetric-scattering approximations, such as relaxation-time, where $w_{\bm k\rightarrow \bm k'} = w_{\bm k'\rightarrow \bm k}$, $f^{(2)}_{\bm k}$ remains even in momentum unless time reversal symmetry is broken. Thus, within this widely used approximation, the longitudinal nonlinear current vanishes.

We show that this restriction is lifted once disorder-induced asymmetric scattering is included. In noncentrosymmetric systems, the scattering probability contains an antisymmetric component satisfying $w_{\bm k\rightarrow \bm k'} \neq w_{\bm k'\rightarrow \bm k}$, while maintaining microscopic reversibility and detailed balance (see Appendix A2). 
Such asymmetric scattering generates a $\bm k$-odd component of $f^{(2)}_{\bm k}$, which combines with the $\bm k$-odd band velocity to produce a finite longitudinal current proportional to $E^2$. Two extrinsic mechanisms generate this asymmetry. Skew scattering produces a directional imbalance in impurity scattering that directly induces a $\bm k$-odd correction to the distribution function, while impurity scattering can also produce a real-space coordinate shift of the electronic wave packet, giving rise to a side-jump velocity and additional corrections to the distribution. Both mechanisms therefore provide microscopic channels for longitudinal nonreciprocal charge transport even in time-reversal-symmetric systems.


\tc{blue}{\textit{Disorder-driven longitudinal nonreciprocity:---}}%
In a two-terminal geometry, we can express the longitudinal electric current up to second-order as, 
\begin{equation}
j_a = \sigma^{\rm eff}_{aa} E_a~~~~{\rm where}~~~~\sigma^{\rm eff}_{aa} (\bm E) \equiv \sigma_{aa} + \sigma_{aaa}E_a. 
\end{equation} 
Here, $\sigma_{aa}$ and $\sigma_{aaa}$ denote the linear and second-order conductivity tensors, whereas $\sigma^{\rm eff}_{aa} (E)$ represent the total effective conductivity. The nonlinear conductivity $\sigma_{aaa}$ controls the magnitude of longitudinal nonreciprocal charge transport. To calculate these conductivities, we use the semiclassical expression for the current, $\bm j = -e \sum_{l} \dot{\bm r}_l f_l$, 
where $l=(n,\bm k)$ labels a Bloch state and $\dot{\bm r}_l$ is the wave-packet velocity.
The nonequilibrium distribution function $f_l$ is obtained from the steady-state Boltzmann equation in a homogeneous field, $
\dot{\bm k}\cdot\nabla_{\bm k}f_l = I_{\rm el}[f_l]$,
with $\hbar\dot{\bm k}=-e\bm E$ and $I_{\rm el}$ the elastic collision integral~\cite{ashcroft1976ssp}.

In real materials, impurity scattering transfers electrons between states $l$ and $l'$ with scattering rate probability $w_{ll'} \equiv w_{l \to l'}$.
In noncentrosymmetric systems, this scattering rate can generally be decomposed into symmetric ($w^S_{ll'}$) and antisymmetric parts ($w^A_{ll'}$),
\begin{equation} \label{eq:SR_comp}
w_{ll'} = w^S_{ll'} + w^A_{ll'},~~
{\rm where}~~
w^{S/A}_{ll'}=\frac{w_{ll'} \pm w_{l'l}}{2}~.
\end{equation}
The symmetric part describes conventional momentum relaxation, whereas the antisymmetric part arises from extrinsic skew scattering and side-jump processes. In addition, extrinsic impurity scattering also induces a real-space shift of the electronic wave packet during a collision, generating a side-jump velocity. 

Including these disorder-induced processes, the electrical conductivity can be expressed as,
\be 
\label{eq:long_cond_Tot}
\sigma_{aaa} = \sigma^{\rm ND}_{aaa} + \sigma^{\rm NSJ}_{aaa} + \sigma^{\rm NSK}_{aaa}~,
\ee 
where $\sigma^{\rm ND}$, $\sigma^{\rm NSJ}$, and $\sigma^{\rm NSK}$ denote the nonlinear Drude, side-jump, and skew-scattering contributions, respectively.
In $\mathcal{T}$-symmetric systems, the linear conductivity solely arises from the Drude contribution ($\sigma_{aa}^{\rm D}$), whereas nonlinear longitudinal conductivity originates from side-jump and skew-scattering contributions, which dominate the longitudinal NCT. 

Solving the Boltzmann equation beyond the relaxation-time approximation (see Sec.~1 of the Supplemental Material (SM)~\cite{NCT_SM} for details), we obtain linear Drude and nonlinear side-jump  and skew scattering conductivities as
\begin{subequations}
\begin{align}
\sigma^{\rm D}_{aa}
&=
-e^2\tau
\sum_l
v_l^a v_l^a
\partial_{\epsilon_l} f_l^0 ,
\\
\sigma^{\rm NSJ}_{aaa}
&=
-\frac{2e^3\tau^2}{\hbar}
\sum_l
v^{\rm sj,a}_l
\partial_a v_l^a
\partial_{\epsilon_l} f_l^0 ,
\\
\sigma^{\rm NSK}_{aaa}
&= -
\frac{e^3\tau^3}{\hbar^2}
\sum_{l,l'}
w^A_{ll'}
\left[
v^a_{l'}\partial_a^2 f_l^0
-
(\partial'_a v^a_{l'})
\partial_a f_l^0
\right].
\end{align}
\end{subequations}
Here,$v_l^a=\hbar^{-1}\partial_{k_a}\epsilon_l$ is the group velocity and $f_l^0$ is the equilibrium Fermi–Dirac distribution.
The side-jump velocity is defined as $\bm v^{\rm sj}_l = \sum_{l'} w^S_{ll'}\,\delta\bm r_{ll'}~,$ where $\delta\bm r_{ll'}$ is the coordinate shift of the wave packet during scattering~\cite{sinitsyn2006coordinate}. Importantly, both $\sigma^{\rm NSJ}$ and $\sigma^{\rm NSK}$ are $\mathcal{T}$-even responses, consistent with microscopic reversibility, allowing them to remain finite in $\mathcal{T}$-symmetric systems. 

Although these contributions originate from disorder, their magnitude is closely linked to the underlying band geometry.
In particular, both the side-jump velocity and the antisymmetric scattering rate are controlled by the Berry curvature of the electronic bands~\cite{varshney2026asymmetric}. We show and discuss the band geometric origin of the coordinate shift and skew-scattering rate explicitly in Appendix A3. Thus, longitudinal nonlinear conductivity offers an independent probe of band geometry, even in systems where the nonlinear Hall response may vanish due to crystalline symmetries. 
Noncentrosymmetric materials with strong Berry curvature provide natural platforms for large disorder-driven longitudinal NCT.

\begin{table}[t!]
\renewcommand{\arraystretch}{1.25} 
    \centering
    \small
    \setlength{\tabcolsep}{4pt}
    \caption{List of all 42 point groups (both nonmagnetic and magnetic) that allow a finite ${\mathcal T}$-even longitudinal conductivity $\sigma_{aaa}$. While our focus is on nonmagnetic systems, the contributions of $\sigma^{\rm NSJ}_{aaa}$ and $\sigma^{\rm NSK}_{aaa}$ in magnetic systems also remain largely unexplored.}
    \begin{tabular}{lll}
    \hline \hline
         $\mathcal{T}$-even $\sigma_{aaa}$ &
         \parbox[t]{0.24\columnwidth}{\raggedright Nonmagnetic point groups} &
         \parbox[t]{0.50\columnwidth}{\raggedright Magnetic point groups} \\
         \hline\hline 
        $\sigma_{xxx}$ &
        \parbox[t]{0.24\columnwidth}{\raggedright $1$, $m$, $3$, $32$, $-6$} &
        \parbox[t]{0.50\columnwidth}{\raggedright $1.1'$, $m.1'$, $m'$, $3.1'$, $32.1'$, $32'$, $-6.1'$, $-6'$} \\
         \hline 
         $\sigma_{yyy}$ &
         \parbox[t]{0.24\columnwidth}{\raggedright $1$, $2$, $3$, $3m$, $-6$, $-6m2$} &
         \parbox[t]{0.50\columnwidth}{\raggedright $1.1'$, $2.1'$, $2'$, $3.1'$, $3m.1'$, $3m'$, $-6.1'$, $-6'$, $-6m2.1'$, $-6m'2'$, $-6'm2'$, $-6'm'2$} \\ \hline 
        $\sigma_{zzz}$ &
        \parbox[t]{0.24\columnwidth}{\raggedright $1$, $m$, $mm2$, $4$, $4mm$, $3$, $3m$, $6$, $6mm$} &
        \parbox[t]{0.50\columnwidth}{\raggedright $1.1'$, $m.1'$, $m'$, $mm2.1'$, $m'm'2$, $m'm2'$, $4.1'$, $4'$, $4mm.1'$, $4m'm'$, $4'm'm$, $3.1'$, $3m.1'$, $3m'$, $6.1'$, $6'$, $6mm.1'$, $6m'm'$, $6'mm'$}
        \\ \hline \hline
    \end{tabular}
    \label{MPGs_tab}
\end{table}


\tc{blue}{\textit{Symmetry constraints on nonreciprocity:---}}
The longitudinal nonlinear conductivity $\sigma_{aaa}$ is strongly constrained by crystal symmetry. Inversion symmetry forces $\sigma_{aaa}$ to vanish, so noncentrosymmetric materials are required. As discussed in Eq.~\eqref{eq:long_cond_Tot}, the nonlinear Drude term $\sigma^{\rm ND}_{aaa}$ is odd under time reversal and therefore requires broken $\mathcal{T}$. In contrast, the disorder-induced side-jump and skew-scattering contributions $\sigma^{\rm NSJ}_{aaa}$ and $\sigma^{\rm NSK}_{aaa}$ are $\mathcal{T}$-even and can remain finite in nonmagnetic systems. Thus, inversion breaking alone is sufficient to allow longitudinal NCT driven by these extrinsic mechanisms.

\begin{figure}[t!]
    \centering
    \includegraphics[width = 0.95\linewidth]{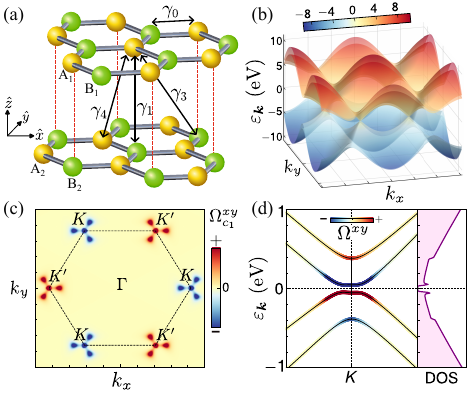}
    \caption{\textbf{Bernal-stacked bilayer graphene (BLG) and its electronic properties.} (a) Crystal structure of BLG showing the AB stacking and the relevant intra- and interlayer hopping processes. (b) Electronic band dispersion of BLG. (c) Berry curvature distribution of the first conduction band in momentum space, exhibiting pronounced hot spots near the $K$ and $K'$ valleys. The dashed hexagon indicates the first Brillouin zone with marked high-symmetry points. (d) Band structure near the $K$ valley, color-coded by the Berry curvature magnitude, together with the density of states. BLG parameters are discussed in Sec.~S2 of the SM~\cite{NCT_SM}, and the onsite potential is $\Delta = 0.05$ eV.} 
    \label{fig2}
\end{figure}

To identify candidate materials that support the $\cal T$-even $\sigma_{aaa}^{\rm NSJ}$ and $\sigma_{aaa}^{\rm NSK}$ contributions, we perform a symmetry analysis across all 122 point groups. Following Neumann’s principle~\cite{Newnham2004properties} and using the Bilbao Crystallographic Server~\cite{Gallego2019bilbao}, we find 42 point groups, including magnetic ones, that permit a nonvanishing $\sigma_{aaa}$. The classification is summarized in Table~\ref{MPGs_tab}. Experimentally relevant noncentrosymmetric classes include $3.1'$, $3m.1'$, $-6.1'$, and $-6'$. Candidate platforms include Weyl semimetals~\cite{Ma2018observation, Kumar2021room}, tellurium~\cite{Cheng2024giant, manuel2024odd}, multilayer graphene  heterostructures~\cite{panhe2022graphene, Huang2022giant, Duan2022giant, Sinha2022berry, Datta2024nonlinear, Ahmed2025detecting}, and Rashba systems~\cite{Lu2024nonlinear}, amongst others.

\tc{blue}{\textit{Large longitudinal nonreciprocity in BLG:---}}
To illustrate the magnitude of the longitudinal NCT response, we consider Bernal-stacked bilayer graphene (BLG), shown in Fig.~\ref{fig2}(a). BLG is a natural platform because its carrier density and inversion asymmetry can be tuned independently in a dual-gate device. A vertical displacement field breaks inversion symmetry while preserving time reversal. Gated BLG preserves $\mathcal{T}$ and $\mathcal{C}_{3z}$. Therefore, the $\mathcal{T}$-odd nonlinear Drude contribution vanishes, while the $\mathcal{T}$-even side-jump and skew-scattering terms are allowed. In addition, BLG retains a mirror symmetry $\mathcal{M}_y$ about the armchair direction (see Fig.~S1 of the SM~\cite{NCT_SM}), which forces $\sigma_{yyy} = 0$. Therefore, the only symmetry-allowed longitudinal NCT response in pristine BLG is $\sigma_{xxx}$, where the $x$ direction is chosen along the zigzag axis.

We follow Refs.~\cite{McCann2013, kuzmenko2009determination, koushik2025planar} for the tight-binding model of AB-stacked BLG that captures the electronic structure, including the low-energy bands near the Dirac points (details in Sec.~S2 of the SM~\cite{NCT_SM}). 
The resulting dispersion is shown in Fig.~\ref{fig2}(b). 
The displacement-field breaks inversion symmetry and generates a finite Berry curvature, concentrated near the $K$ and $K'$ valleys [Fig.~\ref{fig2}(c)]. The band structure around the $K$ valley, weighted by Berry curvature, together with the density of states, is shown in Fig.~\ref{fig2}(d). 
Trigonal warping produces low-energy van Hove singularities near the band edges, giving rise to pronounced peaks in the density of states (see Sec.~S3 of the SM~\cite{NCT_SM} for details). These van Hove singularities lying near the band edges strongly enhance the nonreciprocal response, as we discuss below. 

\begin{figure}[t!]
    \centering
    \includegraphics[width=0.95\linewidth]{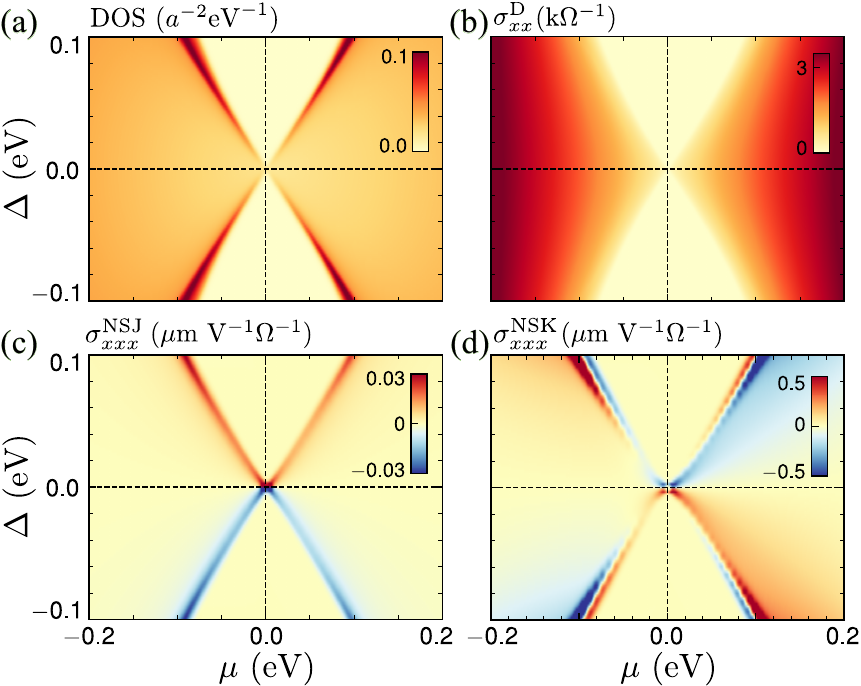}
    \caption{\textbf{Linear and nonlinear conductivities in bilayer graphene in the $\mu$-$\Delta$ plane.}  
    (a) Density of states, showing van Hove singularities near the band edges. 
    (b) Linear Drude conductivity $\sigma_{xx}=\sigma_{yy}$. 
    (c) Nonlinear side-jump contribution $\sigma^{\rm NSJ}_{xxx}$. 
    (d) Nonlinear skew-scattering contribution $\sigma^{\rm NSK}_{xxx}$. 
    We considered $T=20$ K with model parameters identical to Fig.~\ref{fig2}.}
    \label{fig3}
\end{figure}

We next evaluate the linear and nonlinear conductivities as functions of the experimentally controllable chemical potential $\mu$ and displacement-field potential $\Delta$. The results are shown in Fig.~\ref{fig3}. Van Hove singularities near the band edges strongly enhance the nonlinear conductivities, while the linear conductivity remains smooth and nearly symmetric in $\mu$ and $\Delta$. The side-jump contribution $\sigma^{\rm NSJ}_{xxx}$ [Fig.~\ref{fig3}(c)] is symmetric in $\mu$ and antisymmetric in $\Delta$, whereas the skew-scattering contribution $\sigma^{\rm NSK}_{xxx}$ [Fig.~\ref{fig3}(d)] is antisymmetric in both $\mu$ and $\Delta$ and changes sign across the van Hove singularities. For the disorder parameters used here, skew scattering dominates the nonlinear response (Appendix~A3).

\tc{blue}{\textit{Experimental signature of NCT in BLG:---}}
To quantify the magnitude of longitudinal nonreciprocity, we introduce the dimensionless nonreciprocity factor~\cite{Xiao2026lorentz},
\be \label{eq:NR_factor}
\eta = \left\vert \frac{\sigma^{\rm eff}(E_a) - \sigma^{\rm eff}(-E_a)}{\sigma^{\rm eff}(E_a) + \sigma^{\rm eff}(-E_a)}\right  \vert
= \left\vert \frac{\sigma_{aaa} E_a}{\sigma_{aa}} \right\vert
\equiv \left|\frac{V^{2\omega}_{aa}}{V^{\omega}_{aa}}\right|~.
\ee 
Here, $E_a$ is the magnitude of the applied electric field and  
$\eta$ quantifies the asymmetry between forward and reverse currents. In nonlinear transport experiments, the response is detected through the first- and second-harmonic voltages $V^{\omega}_{aa}$ and $V^{2\omega}_{aa}$. The conductivity ratio can be expressed in terms of these voltages as $\sigma_{aaa}E_a/\sigma_{aa} = V^{2\omega}_{aa}/V^{\omega}_{aa}$~\cite{panhe2022graphene} (details in Sec.~S4 of the SM~\cite{NCT_SM}). This allows a direct estimation of the nonreciprocity factor from the measured harmonic voltages. For the asymmetric scattering mechanism, we analyze the disorder dependence of $\eta$  in Sec. S5 of SM~\cite{NCT_SM}. 

\begin{figure}[t!]
    \centering
    \includegraphics[width=0.95\linewidth]{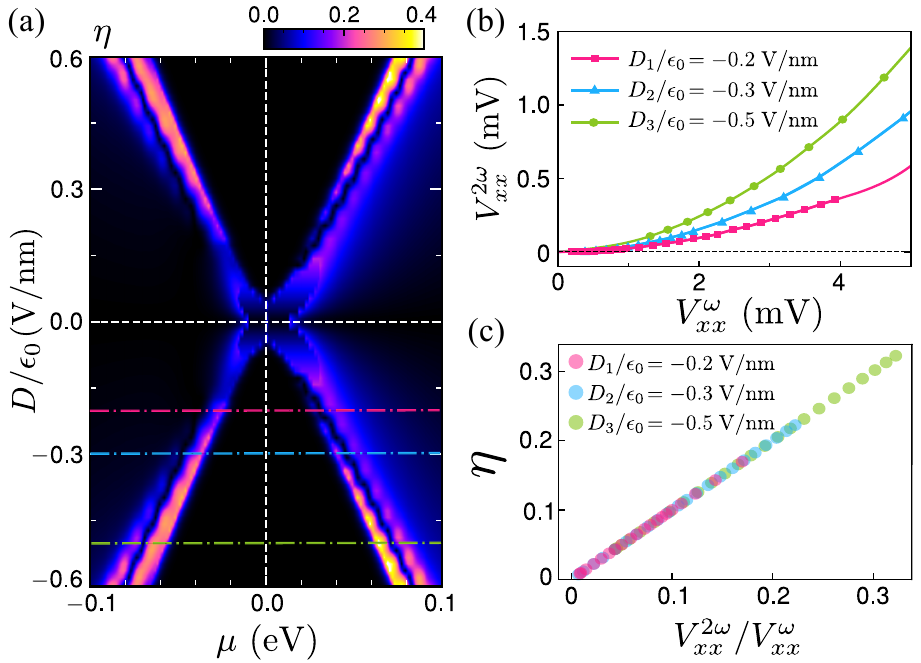}
    \caption{\textbf{Large longitudinal nonreciprocity in gated bilayer graphene.} 
    (a) Calculated nonreciprocity factor $\eta$ as a function of chemical potential $\mu$ and displacement field $D$. Strong nonreciprocity appears near the band edges, where linear conductivity is small and nonlinear conductivity is large. 
    (b) Experimentally measured second-harmonic voltage $V^{2\omega}_{xx}$ versus first-harmonic voltage $V^{\omega}_{xx}$ for different displacement fields $D$, reproduced from Ref.~\cite{Ahmed2025detecting}.
(c) Nonreciprocity factor $\eta$ extracted from the experimental data using Eq.~\eqref{eq:NR_factor}, reaching values above $30\%$ and showing good qualitative agreement with theory. The dashed lines in (a) indicate the corresponding experimental values of $D$. }
    \label{fig4}
\end{figure}

Using the calculated linear and nonlinear conductivities of bilayer graphene, we map the nonreciprocity factor $\eta$ in the $(\mu,\Delta)$ plane in Fig.~\ref{fig4}(a). The response is enhanced near the band edges, where van Hove singularities suppress the linear conductivity and amplify the nonlinear response. In our model, $\eta$ reaches $\sim0.4$ for an electric field $E\sim0.5$~V/mm, within the experimental range. 
The nonreciprocity is also strongly tunable with the displacement field $\Delta$.

To benchmark the disorder-driven mechanism against available experiment, we compare our results with recent measurements of nonlinear longitudinal transport in bilayer graphene~\cite{Ahmed2025detecting}. The experimental signal is measured in the longitudinal $xx$ channel, which is the zigzag direction also used for the theoretical $\sigma_{xxx}$ calculations. Figure~\ref{fig4}(b) shows the measured $V^{2\omega}_{xx}$ as a function of $V^{\omega}_{xx}$ for several displacement fields $D$. Using Eq.~\eqref{eq:NR_factor}, we convert these data into the corresponding nonreciprocity factor and plot the result in Fig.~\ref{fig4}(c). The extracted $\eta$ increases with $|\Delta|$ and reaches $\eta\sim0.3$ for $D/\epsilon_0 =-0.5$~V/nm
\footnote{{The displacement field is related to the onsite potential difference through $\Delta=eDc/\epsilon_0 \epsilon_r$, where $c=0.335$~nm is the interlayer spacing and $\epsilon_r\approx2.6$ is the effective dielectric constant of BLG~\cite{Slizovskiy2021out}. The experimental range $D/\epsilon_0 =[-0.2,-0.3,-0.5]$~V/nm therefore corresponds to $\Delta\approx[-0.02,-0.04,-0.06]$~eV.}}
. The theoretical and experimental results show good qualitative agreement in displacement-field tunability, and order-of-magnitude agreement. In BLG, the large $\eta$ reflects the combined effect of strong Berry curvature and van Hove singularities near the band edges. 

Unlike nonlinear Hall responses, which are governed by Berry-curvature-dipole physics, the present effect appears directly in the longitudinal channel and is controlled by asymmetric scattering. It has received less attention in nonmagnetic systems because experiments often emphasize the transverse nonlinear signal. Since the nonreciprocity discussed here is set by the ratio of nonlinear to linear longitudinal response, measurements that track both of these longitudinal signals should provide a useful test of the mechanism.

\tc{blue}{\textit{Conclusion:---}} 
We have shown that longitudinal nonreciprocal charge transport does not require broken time-reversal symmetry. In noncentrosymmetric conductors, disorder-induced asymmetric scattering can generate a longitudinal current proportional to $E^2$ while remaining consistent with microscopic reversibility and detailed balance.

Our symmetry analysis identifies 42 point groups that allow this response. In gated bilayer graphene, the calculated nonreciprocity is enhanced near the Lifshitz transition and agrees with available experiments in displacement-field tunability, and order of magnitude. More broadly, our results point to noncentrosymmetric graphene heterostructures, tellurium, Weyl semimetals, and Rashba systems as promising platforms for realizing longitudinal NCT.

\tc{blue}{\textit{Acknowledgments}:---}
We acknowledge Tanweer Ahmed, Luis E. Hueso, Kamal Das, Sunit Das, and Sayan Sarkar for many fruitful discussions. H.V. acknowledges the Ministry of Education, Government of India, for financial support through the Prime Minister’s Research Fellowship. A.A. acknowledges funding from the Core Research Grant by ANRF (Sanction No. CRG/2023/007003), Department of Science and Technology, India.

\section{Appendix}

\subsection{A1: Generalized Onsager reciprocity and dissipative nonlinear transport} \label{app1}
Onsager reciprocity in nonmagnetic systems follows from microscopic
reversibility, a consequence of time-reversal symmetry of the Hamiltonian
together with equilibrium statistical mechanics. In linear response,
Onsager reciprocity imposes strict constraints on transport coefficients
and forbids longitudinal NCT in the absence of magnetic order or external magnetic fields.

In a nonlinear response, the second-order current generated by an electric field can be written as
$J_i(\omega_1+\omega_2)
=
\sum_{jk}
\sigma_{ijk}(\omega_1,\omega_2)
E_j(\omega_1)E_k(\omega_2)$.
Since the charge current is odd under time reversal, microscopic reversibility implies the generalized Onsager relation in Matsubara frequency~\cite{morimoto2018nonreciprocal, Nagaosa2024nonreciprocal}: 
$\sigma_{ijk}(i\omega_1,i\omega_2) = - \sigma_{ikj}(-i\omega_2,-i\omega_1)$.
If the dc limit were analytic, taking $\omega_1,\omega_2\to0$ would yield
$\sigma_{ijk}(0,0)=-\sigma_{ijk}(0,0)$,
apparently implying $\sigma_{ijk}(0,0)=0$, and forbidding any second-order
dc response in a time-reversal-symmetric system.

This conclusion does not hold once momentum relaxation and dissipation are included. The dc nonlinear response corresponds to a nonequilibrium steady state that necessarily requires dissipation, which is encoded in the collision integral of the Boltzmann equation. As a result, the nonlinear conductivity develops nonanalytic structures of the form
\begin{equation}
\sigma_{ijk}(\omega_1,\omega_2) \sim \frac{1}{\omega_1+i/\tau},
\quad
\frac{1}{\omega_2+i/\tau},
\quad
\frac{1}{\omega_1+\omega_2+i/\tau},
\end{equation}
and their products. After analytic continuation
$i\omega\rightarrow\omega+i0^+$, these generate relaxation poles at $\omega_i=0$.
The limits $\omega_1\to0$ and $\omega_2\to0$ therefore do not commute,
and the dc response cannot be obtained by a smooth analytic continuation
from Matsubara frequencies.

Physically, these poles encode the relaxation dynamics that stabilize
the nonequilibrium steady state. The emergence of such nonanalytic
structures is a generic feature of steady-state transport, where
relaxation processes balance the external driving field. Consequently,
dissipative nonlinear dc responses are not constrained by equilibrium
Onsager reciprocity relations. Longitudinal NCT 
proportional to $E^2$ is therefore allowed in time-reversal-symmetric systems when inversion symmetry is broken, and the response is intrinsically dissipative.

\subsection{A2: Microscopic reversibility, detailed balance, and asymmetric scattering} \label{app2}

Understanding longitudinal NCT in time-reversal-symmetric systems requires distinguishing between \emph{microscopic reversibility} and \emph{detailed balance}. These concepts are closely related but not identical.
If the Hamiltonian preserves time-reversal symmetry ($\mathcal T$), the scattering probability has to satisfy the condition of microscopic reversibility, $w_{\bm k \to \bm k'} = w_{-\bm k' \to -\bm k}$. This relation states that the transition probability for a $\bm k\!\to\!\bm k'$ scattering is equal to that of the time-reversed scattering process $-\bm k'\!\to\!-\bm k$. 

In many simplified treatments, such as the relaxation-time approximation, we further assume symmetric scattering with $w_{\bm k \to \bm k'} = w_{\bm k' \to \bm k}$. This enforces equality between forward and backward processes. However, this is a stronger constraint than microscopic reversibility, and it does not hold in inversion-broken systems. In such materials, the scattering probability can contain an antisymmetric component that gives rise to skew scattering. In general, the scattering rate can be decomposed into symmetric and antisymmetric parts as shown in Eq.~\eqref{eq:SR_comp}. The antisymmetric part produces skew scattering and is essential for generating nonlinear nonreciprocal transport. The presence of such asymmetric scattering does not violate time-reversal symmetry because the microscopic reversibility condition remains satisfied. More importantly, we show below that such antisymmetric scattering terms also maintain detailed balance.

The principle of detailed balance follows from the requirement that the Boltzmann collision integral vanishes in equilibrium. For elastic impurity scattering, the collision integral takes the form
\begin{equation}
I[f(\bm k)]
=
\sum_{\bm k'}
\left[
w_{\bm k' \to \bm k} f_{\bm k'}
-
w_{\bm k \to \bm k'} f_{\bm k}
\right].
\label{eq:collision_integral}
\end{equation}
In equilibrium, the distribution function is the Fermi–Dirac function $f^{(0)}(\varepsilon_{\bm k})$. For elastic scattering we have $\varepsilon_{\bm k}=\varepsilon_{\bm k'}$, which implies $f^{(0)}_{\bm k}=f^{(0)}_{\bm k'}$. Substituting this into Eq.~(\ref{eq:collision_integral}) gives
\begin{equation}
I[f^{(0)}_{\bm k}]
=
f^{(0)}_{\bm k}
\sum_{\bm k'}
\left[
w_{\bm k' \to \bm k}
-
w_{\bm k \to \bm k'}
\right].
\end{equation}
For equilibrium to be maintained, the collision integral must vanish, which requires $\sum_{\bm k'} w_{\bm k' \to \bm k} = \sum_{\bm k'} w_{\bm k \to \bm k'}$. This is the condition of detailed balance, which states that the total probability of scattering into a state equals the total probability of scattering out of that state. The symmetric part of the scattering rate always satisfies this detailed balance relation. Additionally, we can use the microscopic reversibility relation,  $w_{\bm k \to \bm k'} = w_{-\bm k' \to -\bm k}$, to show that the antisymmetric scattering rate satisfies $\sum_{\bm k'} w^A_{\bm k' \to \bm k}=
\sum_{\bm k'} w^A_{\bm k \to \bm k'} = 0$.
This ensures that the total inflow and outflow scattering probabilities are equal, so that $I[f^{(0)}]=0$.

Thus, even when the scattering probability contains an antisymmetric component responsible for skew scattering, equilibrium is preserved and no current flows. Asymmetric scattering, therefore, remains fully consistent with time-reversal symmetry and detailed balance.

\subsection{A3: Band-geometric origin of extrinsic contributions} \label{app3}
The side-jump velocity and the antisymmetric (skew) scattering rate are the key microscopic mechanisms responsible for longitudinal NCT in time-reversal-symmetric systems. Both quantities originate from overlaps between Bloch states at different crystal momenta. In particular, the side-jump velocity explicitly depends on the intraband Berry connection through the coordinate shift of the electronic wave packet. This observation naturally suggests that even disorder-driven responses can encode geometric information of the Bloch bands. To explicitly analyze this connection, we introduce a model for the impurity potential.

\subsubsection{Impurity potential modeling}
To evaluate extrinsic responses associated with side-jump and skew-scattering processes, we consider randomly distributed short-range static scatterers following Refs.~\cite{du2019disorder, papaj2021enhanced, ma2023anomalous, guo2024extrinsic, ma2025quantum, varshney2026asymmetric}. We consider short-range  impurities and model them using a Dirac-delta potential,
\be
V_{\rm imp}({\bm r}) = \sum_j V_j \delta({\bm r} - {\bm R}_j)~.
\ee
Here, $V_j$ denotes the impurity strength at position ${\bm R}_j$, and the summation runs over all impurity sites. For this potential, the matrix element in the Bloch basis takes the form
$V_{ll'} \equiv V^0_{\kb,\kb'} \langle u_l | u_{l'} \rangle$, 
where $V^0_{\kb,\kb'} = \sum_j V_j e^{i(\kb' - \kb)\cdot {\bm R}_j}$ is the Fourier transform of the impurity potential. Therefore, the disorder averages of different orders of $V_{ll'}$ required for side-jump velocity and antisymmetric scattering rate, can be written as 
\be
\begin{aligned}
\langle V_{ll'} V_{l'l} \rangle_{\rm dis}
&=
\langle V^0_{\kb \kb'} V^0_{\kb' \kb} \rangle_{\rm dis}
\langle u_l | u_{l'} \rangle
\langle u_{l'} | u_l \rangle ~,
\\
\langle V_{ll''} V_{l''l'} V_{l'l} \rangle_{\rm dis}
&=
\langle V^0_{\kb \kb''} V^0_{\kb'' \kb'} V^0_{\kb' \kb} \rangle_{\rm dis} \times \\
&~ \langle u_l | u_{l''} \rangle
\langle u_{l''} | u_{l'} \rangle
\langle u_{l'} | u_l \rangle ~,
\\
\langle V_{ll''} V_{l''l'} V_{l'l'''} V_{l'''l} \rangle_{\rm dis}
&=
\langle V^0_{\kb \kb''} V^0_{\kb'' \kb'} V^0_{\kb' \kb'''} V^0_{\kb''' \kb} \rangle_{\rm dis} \\
& \times 
\langle u_l | u_{l''} \rangle
\langle u_{l''} | u_{l'} \rangle
\langle u_{l'} | u_{l'''} \rangle
\langle u_{l'''} | u_l \rangle ~.
\end{aligned} \nn
\ee
For disorder averaging, we simplify 
\be
\begin{aligned}
\langle V^0_{\kb \kb'} V^0_{\kb'\kb} \rangle_{\rm dis} &= n_i V_0^2~, \\
\langle V^0_{\kb \kb''} V^0_{\kb'' \kb'} V^0_{\kb' \kb} \rangle_{\rm dis} &= n_i V_1^3~, \\
\langle V^0_{\kb \kb''} V^0_{\kb'' \kb'} V^0_{\kb' \kb'''} V^0_{\kb''' \kb} \rangle_{\rm dis} &= n_i^2 V_0^4~.
\end{aligned}
\nn
\ee
Here, $n_i$ denotes the impurity concentration, while $V_0$ and $V_1$ represent the zeroth and first moments of the impurity potential.

For numerical calculations, we adopt realistic disorder parameters following Ref.~\cite{ma2023anomalous}: $n_i \approx 10^{10}\,{\rm cm^{-2}}$ and $V_0 = 6.2 \times 10^{-13}\,{\rm eV\,cm^2}$. These values correspond to a symmetric scattering time $\tau \approx 10^{-13}\,{\rm s}$ ($\tau \approx 0.1$ ps), consistent with previous studies~\cite{papaj2021enhanced, Xiao2026lorentz}. To obtain comparable magnitudes of the nonlinear skew-scattering contributions originating from third- and fourth-order processes, we set $V_1 = 0.5 V_0$. Varying $V_1/V_0$ changes the relative magnitude of the skew-scattering channels but does not affect the symmetry analysis; the disorder scaling of $\eta$ is discussed in Sec.~S5 of the SM~\cite{NCT_SM}.

\subsubsection{Berry curvature dependence of side-jump velocity and antisymmetric scattering rate}

Following the analysis of Ref.~\cite{varshney2026asymmetric}, compact expressions for the side-jump velocity and skew-scattering rates can be obtained in the weak-impurity-potential limit, where both quantities acquire a direct dependence on the Berry curvature. For time-reversal-symmetric systems, the extrinsic side-jump velocity and the third- and fourth-order skew-scattering rates are given by
\begin{center}
$\begin{array}{rcl}
    {\bm v}^{\rm sj}_{n, \kb} & = & \dfrac{2\pi}{\hbar} n_i V_0^2 \mathcal{D}_n(\e_n^\kb)  [ \kb \times {\bm \Omega}_n] ~, \\
    w^{(3), A}_{n, \kb\kb' } & = & \dfrac{2\pi^2}{\hbar} n_i V_1^3 \mathcal{D}_n(\e^\kb_n) [(\kb \times \kb') \cdot {\bm \Omega}_n]~ \delta(\e^\kb_n  - \e^{\kb'}_n)~, \\
    w^{(4), A}_{n, \kb\kb' } & = & \dfrac{2\pi^2}{\hbar} n^2_i V_0^4 \mathcal{D}_n(\e^\kb_n) [(\kb \times \kb') \cdot \tilde{\bm \Omega}_n]~ \delta(\e^\kb_n  - \e^{\kb'}_n).
\end{array}$
\end{center}
Here $\bm{\Omega}_n$ denotes the Berry curvature of the $n$th band, while $\tilde{\bm \Omega}_n = \sum_{n'} {\bm \Omega}_{nn'}/(\e^\kb_n - \e^{\kb}_{n'})$ represents the energy-normalized Berry curvature. The quantity $\mathcal{D}_n(\e^\kb_n) = \sum_{\kb''} \delta(\e^\kb_n - \e^{\kb''}_n)$ is the density of states of the $n$th band evaluated at energy $\e^\kb_n$. 

These gauge-invariant expressions show that both the side-jump velocity and the antisymmetric scattering rate are fundamentally controlled by the Berry curvature of the Bloch bands, which peak near the band edges. Consequently, even though these mechanisms originate from impurity scattering, their magnitude and structure are directly governed by the underlying Berry curvature of the bands. 

\bibliography{References}

\end{document}